\documentclass[%
 reprint,
nofootinbib,
 amsmath,amssymb,
 aps,prl
]{revtex4-1}

\usepackage{graphicx}
\usepackage{dcolumn}
\usepackage{bm}
\usepackage{hyperref}
\usepackage{color}
\usepackage{comment}
\usepackage{slashed}
\usepackage{simplewick}

\definecolor{deepblue}{rgb}{0.2,0.2,0.8}

\definecolor{deepred}{rgb}{0.8,0.2,0.2}

\newcommand{\vect}[1]{\boldsymbol{\mathbf{#1}}}

\newcommand{\dd}{{\rm d}}

\newcommand{\eV}{{\, {\rm eV}}}
\newcommand{\keV}{{\, {\rm keV}}}

\newcommand{\GeV}{{\, {\rm GeV}}}

\newcommand{\LL}{{\mathcal{L}}}
\newcommand{\OO}{{\mathcal{O}}}

\begin{document}


\title{Dark Photons in the Solar Basin}

\author{Robert Lasenby}
 \email{rlasenby@stanford.edu}
 \affiliation{Stanford Institute for Theoretical Physics, Stanford University, Stanford, CA 94305, USA}

\author{Ken Van Tilburg}
 \email{kvt@kitp.ucsb.edu}
\affiliation{Kavli Institute for Theoretical Physics, University of California, Santa Barbara, California 93106, USA}

\date{\today}

\begin{abstract}
	Production of dark photons inside the Sun forms
	the basis for the most sensitive probes of such
	particles over a wide mass range. A small fraction of
	dark photons is emitted into gravitationally bound
	orbits, building up a ``Solar basin'' population that
	survives for astrophysically long times.
	We show that this population could lead
	to signals in existing and proposed
	dark matter detection experiments, opening
	up significant new parameter space independent
	of whether dark photons make up the dark matter.
	Even with conservative assumptions, results from current dark matter experiments already constrain new parameter space;
	with fiducial assumptions, a Solar basin population
	of dark photons could be responsible for excess events
	seen in XENON1T. 
	Future low-threshold experiments
	could be sensitive to these Solar-System-bound dark photons down to sub-eV
	masses, at couplings orders of magnitude below current constraints.
\end{abstract}

\maketitle


\section{Introduction}

The large temperatures, densities, and volumes of stars make them efficient emitters of weakly-coupled new particles, prevalent in theories beyond the Standard Model (SM). For sufficiently small couplings, the vast majority of such particles will free-stream out to infinity, as stellar neutrinos do. This flux removes energy from the star, affecting its structure and evolution. Such effects---especially in hotter, denser stellar cores, such as those of horizontal branch (HB) stars, red giants (RGs), and white dwarfs---are leading probes of many kinds of new
particle candidates~\cite{Raffelt_1996, An_2013,Redondo_2013,viaux2013neutrino, ayala2014revisiting,Bertolami:2014wua,hardy2017stellar}.

In addition to this unbound flux, a small
proportion of the new particles will be
produced at sub-escape velocities, and will
thus enter bound orbits. 
This was pointed out by Ref.~\cite{Hannestad_2002}, in
the context of Kaluza-Klein emission
from supernovae. For long-lived stars, 
this ``stellar basin" population can build
up over the lifetime of the star, potentially compensating
for its small production rate. The decays of these particles could affect the stellar atmosphere or lead to signatures in Earth-based experiments~\cite{DiLella_2003,Morgan_2005,oka2017search}.
Alternatively, bound particles in Earth-crossing orbits could
be absorbed in laboratory experiments originally
designed to detect dark matter (DM)~\cite{tilburg2020stellar}.

For many new candidate particles, the unbound flux from denser stars imposes stronger coupling constraints
than the unbound flux from the Sun. Compatibility with these stellar cooling constraints restricts the parameter space in which the Solar basin density could be detected, even when it is larger than the unbound density. 
With optimistic assumptions about the survival time of bound orbits,  current and near-future experiments are just beginning to probe new parameter space of axions with couplings to electrons and masses within an order of magnitude around the Solar core temperature, as discussed in Ref.~\cite{tilburg2020stellar}. Observability prospects are likely less favorable for other axion-like couplings and for scalar couplings.

New vector particles can exhibit quite a different behavior. The simplest example is a massive ``dark photon" that couples directly to the usual electromagnetic current, albeit with a smaller coupling. At dark photon masses below the plasma frequency in stellar cores, in-medium mixing with SM plasma oscillations can suppress the production rate of dark photons~\cite{An_2013, Redondo_2013,hardy2017stellar}. Production in the Sun can thus provide stronger constraints than production in denser stars, where plasma frequencies are higher.
Moreover, since these constraints become weaker at smaller dark photon masses due to a well-known decoupling effect~\cite{An_2013}, they leave more room for small-scale, low-threshold experiments to have promising discovery potential, both to a dark matter abundance, and to the Solar flux.

The medium dependence of production rates can also enhance the relative importance of bound versus unbound emission from the Sun itself. For a light dark photon, the outer layers of the Sun can be responsible for the bulk of the dark photon luminosity, because the plasma frequency is lower there. This results in a emission spectrum that is much softer than that of axions or scalars, whose emission is dominated by the stellar core. As we will describe in this work, the softer spectrum is responsible for an increased bound-orbit emission fraction.

In this paper, we will show how the combination of
these effects means that the ``Solar basin" population
of dark photons can open up significant new parameter
space for existing and proposed dark matter detection experiments. We stress that the signals described here do not rely on the dark photon comprising some or all of the dark matter. New limits and future prospects in the dark-photon parameter space are depicted in Fig.~\ref{fig:epsilon}. Our main points are:
\begin{itemize}
\item Even with conservative assumptions about bound-orbit survival, the search of Ref.~\cite{aprile2019light} constrains new parameter space.
\item With more optimistic assumptions about gravitation ejection lifetimes motivated by simulations and secular perturbation theory arguments, the Solar basin population could account for the few-keV excess seen in Ref.~\cite{aprile2020observation}. If these long ejection times are confirmed by simulations, the searches of Refs.~\cite{aprile2019light,aprile2020observation} would set the most stringent DM-independent limit on dark-photon kinetic mixing over much of the mass region $12\,\mathrm{eV} \lesssim m \lesssim 5.4 \, \mathrm{keV}$.
\item Future low-threshold DM detection experiments will be sensitive to couplings orders of magnitude smaller than those currently constrained, and down to sub-electronvolt masses. 
\end{itemize}
In addition to dark photons strictly defined,
our results apply to other light vectors which couple to electrons, such as vectors coupled to baryon-minus-lepton number ($B-L$). For vectors coupled to non-conserved currents, higher-energy processes may place strong constraints on their coupling~\cite{Dror_2017_prl,Dror_2017,Dror_2020}.


\section{Dark Photon Production}

A dark photon $A'$ of mass $m$ that interacts with the SM via a kinetic mixing $\epsilon$, has the Lagrangian:
\begin{align}
\mathcal{L} &\supset -\frac{1}{4} F_{\mu \nu} F^{\mu \nu}- \frac{1}{4} F'_{\mu \nu} F^{\prime \mu \nu}  + \frac{\epsilon}{2} F_{\mu\nu}F^{\prime \mu \nu} \nonumber\\
&\phantom{\supset} + \frac{m^2}{2} A'_\mu A^{\prime \mu} + A_\mu J^\mu_\mathrm{EM}.
\end{align}
We use $+---$ metric signature and natural units with $c = \hbar = k_B = 1$ unless otherwise stated.
Under a field redefinition, this Lagrangian is equivalent to a massive vector that couples to the electromagnetic (EM) current density $\LL \supset \epsilon A'_\mu J^\mu_{\rm EM}$.

In a thermal medium, the propagation eigenstates are no longer
simply the massless SM photon $A$ and the dark photon $A'$.
However, when $\epsilon$ is small, they are only small
perturbations of the massive dark photon and the usual SM
plasma oscillations. The self-energy (in the sense of 
thermal field theory~\cite{Bellac:2011kqa})
for the weakly-coupled, dark-photon-like mode in a
uniform medium is~\cite{Redondo_2013,hardy2017stellar}
\begin{equation}
	\Pi' = \epsilon^2 \Pi + \frac{(\epsilon \Pi)^2}{m^2 - \Pi}
	+ \OO(\epsilon^4)
	= \frac{\epsilon^2 m^2 \Pi}{m^2 - \Pi} + \OO(\epsilon^4),
	\label{eqpi2}
\end{equation}
where we assume that the mode is close to on-shell ($\omega^2 - \vect{k}^2 \simeq m^2$),
and $\Pi$ is the self-energy of the SM photon.
Since the medium is assumed to be uniform, we have suppressed
the polarization indices. 

In $(A,A')$ field space, the weakly-coupled
propagation eigenstate is $(\frac{\epsilon \Pi}{m^2 - \Pi}, 1) + \OO(\epsilon^2)$. If the properties of the Solar medium change slowly compared
to the wavelength of the mode, then a propagating mode will
adiabatically track the local propagation eigenstate,
until in free space it is simply the massive $A'$ mode.
A weakly-coupled mode emitted in the interior of
a star will propagate outwards and thus escape as a massive dark photon, if it has sufficient kinetic energy to do so.

The in-medium production and absorption rates for the weakly-coupled mode
are set by the imaginary part of the self-energy: $\Pi'_i \equiv \mathrm{Im} \lbrace \Pi'\rbrace$. Specifically,
the rate of change of the phase space density $f = (2\pi)^3 \frac{\dd N}{\dd^3 \vect{k} \, \dd^3 R}$ of particles (in each polarization state) is 
\begin{align}
	\dot f &= \Gamma_{\rm prod}' (1 + f) - \Gamma_{\rm abs}' f = \frac{-\Pi'_i}{\omega} \left(\frac{1}{e^{\omega/T} - 1} - f\right).
	\label{eqdotf}
\end{align}
Using the transverse polarization vectors,
we find that 
\begin{align}
\Gamma_{A',\mathrm{prod}}^{(\mathrm{T})} = \frac{\epsilon^2 m^4}{e^{\omega/T}-1} \frac{ \Gamma_{A}^{(\mathrm{T})}}{\big(m^2-\omega_p^2\big)^2 + \big(\omega \Gamma_{A}^{(\mathrm{T})}\big)^2 }
\end{align}
where $\Gamma_A^{\rm (T)} = - \Pi_i^{\rm (T)}/\omega$ is the in-medium transverse
photon width (evaluated on the mass shell $(\omega,\vect{k})$ of the dark photon). The ``plasma frequency" (squared) is the real part of the self-energy $\omega_p^2 = \mathrm{Re} \lbrace \Pi^{\rm (T)} \rbrace$.
The general expression for the longitudinal polarization is more complicated,
but in a uniform, non-relativistic medium, can be simplified to
\begin{align}
	\Gamma_{A',\mathrm{prod}}^{(\mathrm{L})} \simeq \frac{\epsilon^2 m^2 \omega^2}{e^{\omega/T}-1} \frac{ \Gamma_{A}^{(\mathrm{L})}}{\big(\omega^2-\omega_{p}^2\big)^2 + \big(\omega \Gamma_{A}^{(\mathrm{L})}\big)^2 }
\end{align}
where $\Gamma_A^{\rm (L)}$ is the physical width of longitudinal plasma
excitations, with $\Gamma_A^{\rm (L)} \simeq \Gamma_A^{\rm (T)} \equiv \Gamma_A$.

In a non-relativistic electron-ion plasma, such as the interior
of the Sun, the plasma frequency is dominated by the electron
density, yielding $\omega_p^2 \simeq e^2 n_e / m_e$.
The main contributions to the plasma oscillation width $\Gamma_A$ are
electron-ion bremsstrahlung and Thomson scattering~\cite{Redondo_2013,Brussaard_1962}:
\begin{align}
	\Gamma_A &\simeq \frac{16 \pi^2 \alpha^3}{3 m_e^2 \omega^3}
	\sqrt{\frac{2 \pi m_e}{3 T}} (1 - e^{-\omega/T}) n_e \sum_i n_i Z_i^2 
	\bar g_i (\omega,T)	\nonumber \\
    &\phantom{\simeq} + \frac{8 \pi \alpha^2 n_e}{3 m_e^2} \sqrt{ 1- \frac{\omega_p^2}{\omega^2}}
\end{align}
The Thomson scattering expression in the second line is valid for $\omega \ge \omega_p$, since $\omega_p$ is the minimum energy that an in-medium photon can have. 
In the bremsstrahlung expression on the first line, the sum is over the different ion
species, with number densities $n_i$ and charges $Z_i$. For each of these, $\bar g_i(\omega,T)$ is the thermally-averaged
Gaunt factor~\cite{Brussaard_1962}. In the Born approximation,
\begin{equation}
	\bar g = \frac{\sqrt{3}}{\pi} e^{\omega/(2 T)} K_0(\omega/(2 T)),
\end{equation}
valid for $\omega \ll T$ (though $\bar g$ is order unit throughout), matching the expression from Ref.~\cite{Redondo_2013}.


\section{Solar production rates}

\begin{figure}
\includegraphics[width = 0.5 \textwidth]{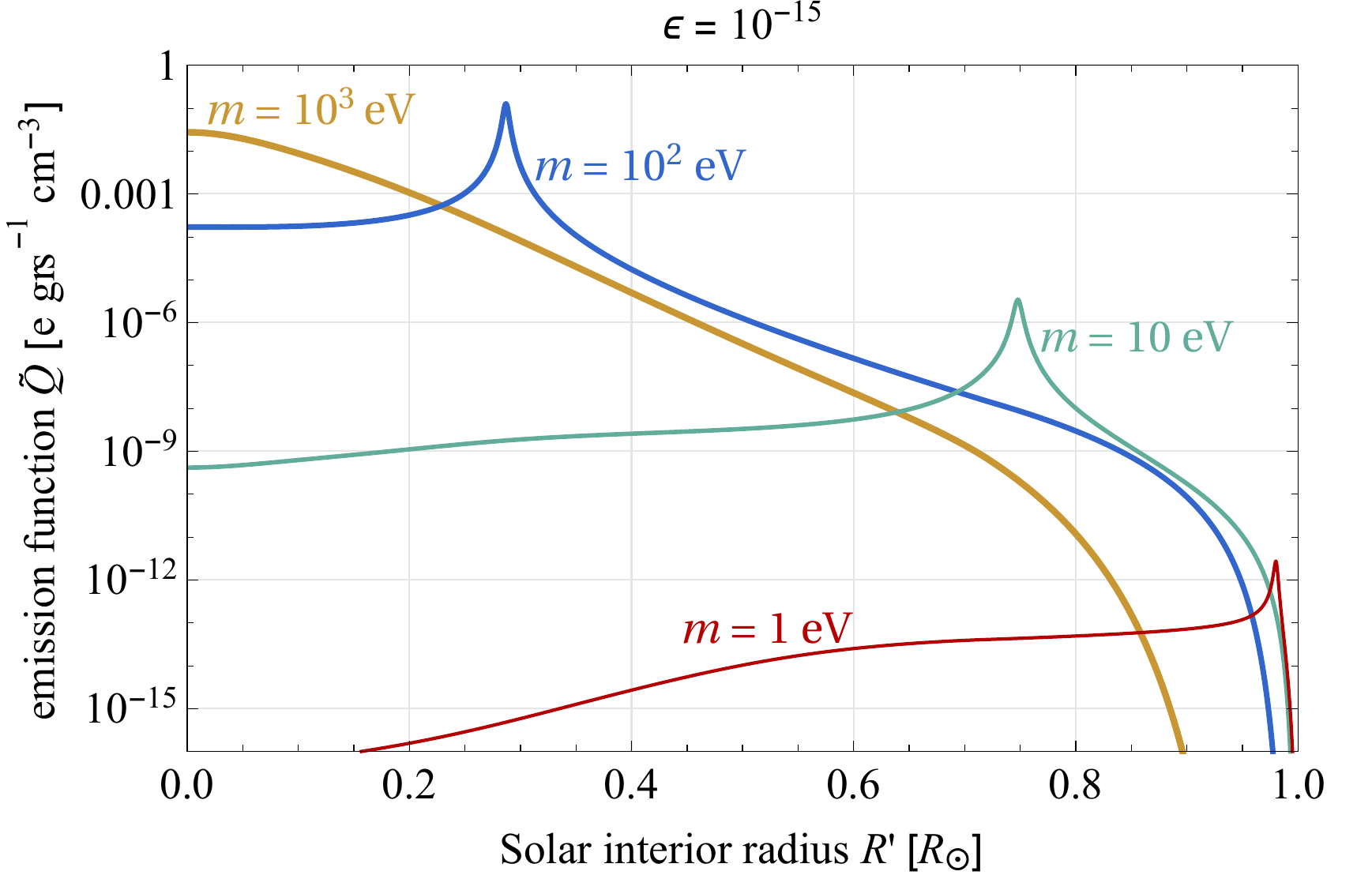}
\caption{Specific energy loss function $\widetilde{Q}$ of Eq.~\ref{eq:Qtilde} for dark photon emission into bound orbits as a function of radius $R'$ within the Sun, for four dark photon masses and a benchmark kinetic mixing $\epsilon = 10^{-15}$. 
For dark photon masses $m$ above the plasma frequency $\max\lbrace\omega_\mathrm{p}\rbrace \approx 290 \eV$ in the Solar core, plasma effects are insignificant. For lower masses, there is a radius within the Sun at which $\omega_\mathrm{p}(R') = m$ and emission is resonantly enhanced.} \label{fig:pbound}
\end{figure}

The energy loss rate per unit volume to weakly-coupled modes is:
\begin{equation}
	Q = \frac{\dd E}{\dd V \dd t} = 
	\int \frac{\dd^3 \vect{k}}{(2\pi)^3} \omega 
	\left(\Gamma^{\rm (L)}_{A',\rm prod}
	+ 2 \Gamma^{\rm (T)}_{A',\rm prod}\right).
\end{equation}
To calculate the total energy flux, we
can simply integrate this expression over the
Solar volume. 
Demanding that the energy flux is less than 10\% of the solar luminosity $L_\odot$, i.e.~$\int \dd^3R' \, Q \lesssim 0.1 L_\odot$, yields a Solar cooling constraint on $\epsilon$~\cite{An_2013,Redondo_2013}.

For non-relativistic production---relevant for emission into bound orbits---the transverse
and longitudinal production rates are approximately equal $\Gamma_{A',\rm prod} \equiv \Gamma_{A',\rm prod}^{\rm (T)} \simeq 
	\Gamma_{A',\rm prod}^{\rm (L)}$, and take the form:
\begin{equation}
	\Gamma_{A',\rm prod}\simeq 
	\frac{\epsilon^2 m^4}{e^{m/T}-1} \frac{\Gamma_A}{(m^2 - \omega_p^2)^2 + 
m^2 \Gamma_A^2}.
\end{equation}
Following Ref.~\cite{tilburg2020stellar}, it is convenient to parametrize the phase space in terms
of the kinetic energy per unit mass: $\tilde \omega_k \simeq k^2/(2 m^2)$.
Since $\frac{\dd^3 \vect{k}}{(2\pi)^3} \simeq \frac{m^3}{2\pi^2} \sqrt{2 \tilde \omega_k} \dd\tilde \omega_k$, we have 
$\frac{\dd Q}{\dd \tilde \omega_k} \simeq \widetilde Q \sqrt{\tilde \omega_k}$, where
\begin{equation}
	\widetilde Q = \frac{3 \epsilon^2 m^8}{\sqrt{2} \pi^2} \frac{1}{e^{m/T}-1}
	\frac{\Gamma_A}{(m^2 - \omega_p^2)^2 + m^2 \Gamma_A^2}. \label{eq:Qtilde}
\end{equation}
We can use this expression to compute the emission
rate into any part of bound-orbit phase space.
A particle must have $\tilde \omega_k \simeq |\Phi|$
at its emission site to reach a radius $R = 1\, \mathrm{AU}$, since the gravitational potential $\Phi$ 
inside the Sun is much deeper than at 1 AU. 
The rate of change 
of the bound state density at radius $R \gg R_\odot$ is
approximately~\cite{tilburg2020stellar}:
\begin{equation}
	\dot \rho_\mathrm{b}(R) \simeq \frac{3}{16 \pi} \frac{G M_\odot}{R^4} \int \dd^3 R'\,
	\widetilde Q(R') \sqrt{|\Phi(R')|},
	\label{eq:rhodot}
\end{equation}
where $R \approx 1\, \mathrm{AU}$ for Earth's radius, $G$ is Newton's gravitational constant, $M_\odot$ is the mass of the Sun. The basin at Earth's location is initially composed of an ensemble of particles on very eccentric orbits, since they are injected on Sun-crossing orbits.

The behaviour of the integrand in Eq.~\ref{eq:rhodot}
depends on the dark photon mass $m$. At masses above the maximum plasma frequency in the Sun,
\begin{equation}
	\widetilde Q \simeq \frac{3 \epsilon^2 m^4}{\sqrt{2} \pi^2} \frac{1}{e^{m/T}-1}
	\Gamma_A, \quad \left(m \gg \omega_p\right)
\end{equation}
and the production rate is dominated by the Solar core,
where $\Gamma_A$ is largest. 
Conversely, if there is some radius $R'_\mathrm{res}$ inside the Sun where resonance occurs
$\omega_p(R'_\mathrm{res}) = m$, then:
\begin{equation}
	\widetilde Q \simeq \frac{3 \epsilon^2 m^4}{\sqrt{2} \pi^2} \frac{1}{e^{m/T_\odot^\mathrm{res}}-1}
	\frac{m^2}{\Gamma_A}, \quad \left(| m - \omega_p | \lesssim \Gamma_A\right)
\end{equation}
at this resonance radius, where the Solar temperature is $T_\odot^\mathrm{res}$.
If $\Gamma_A$ is significantly smaller than $\omega_p$ so that the resonance is narrow, then the
integral over radii in Eq.~\ref{eq:rhodot} is dominated by this narrow range, giving:
\begin{align}
	\dot \rho_\mathrm{b}(R)
	&\simeq \frac{9}{8 \sqrt{2} \pi} \frac{G M_\odot}{R^4} 
	\epsilon^2 m^6 \frac{R_\mathrm{res}'^2 \sqrt{|\Phi(R'_\mathrm{res})|}}{e^{m/T_\odot^\mathrm{res}}-1} \left|\frac{\partial \omega_p}{\partial R'}\right|^{-1}_{R'_\mathrm{res}}. \label{eq:rhodotres}
\end{align}
This expression is independent of $\Gamma_A$, and not suppressed by the fine structure
constant $\alpha$. This is analogous to resonant unbound emission,
as calculated in Refs.~\cite{An_2013,Redondo_2013}.

Figure~\ref{fig:pbound} illustrates these behaviors, showing
the bound energy emission function of Eq.~\ref{eq:Qtilde} as a function of radius for different dark photon masses. The maximum plasma frequency $\omega_{p} \simeq 290 \eV$ is attained in the Solar core; for dark photon masses exceeding this value,
emission from the core dominates. For smaller masses,
there is a resonant feature at the appropriate radius. The integrated bound emission is generally dominated by the resonance expression of Eq.~\ref{eq:rhodotres}, though over the mass range $0.1\,\mathrm{eV} \lesssim m \lesssim 3\,\mathrm{eV}$ the resonance is fractionally less narrow (visible in the red curve of Fig.~\ref{fig:pbound}), and the full expressions of Eqs.~\ref{eq:Qtilde} and \ref{eq:rhodot} should be used.

We obtain the basin energy density injection rate $\dot\rho_\mathrm{b}(R)$ plotted in Fig.~\ref{fig:rho} by integrating over the Solar volume with the Solar model of Ref.~\cite{vinyoles2017new}.
The blue curves represent the time-integrated density $\rho_\mathrm{b} = \dot{\rho}_\mathrm{b} \tau$ for three different values of $\tau$ and a fixed value of $\epsilon = 10^{-16}$.
Since unbound emission (thin blue dotted line) is
also suppressed at smaller $m$, the energy density of the Solar basin can compete with that of the unbound emission down to small masses. This is in contrast to the expected behavior for many other particles, such as axions coupled to electrons, for which the bound-to-unbound energy density ratio rapidly decouples as $\propto m^4$ at low masses~\cite{tilburg2020stellar}.

\section{Dark photon basin evolution}

\begin{figure}
\includegraphics[width = 0.5 \textwidth]{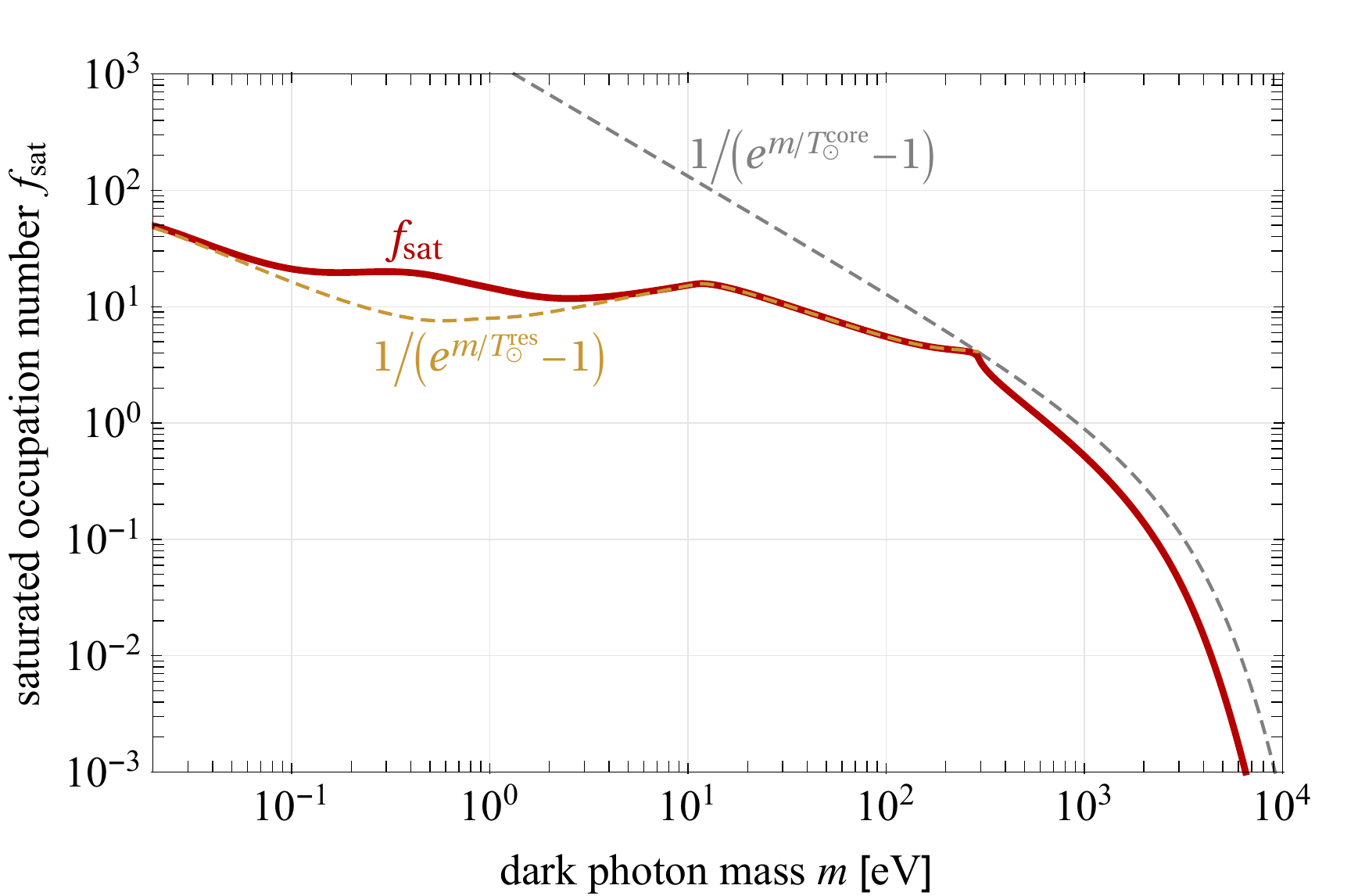}
\caption{Saturated occupation number $f_\mathrm{sat}$ as a function of dark photon mass $m$, plotted in thick red. For low masses $m \lesssim 0.3\,\mathrm{keV}$, dark photon production occurs predominantly in a thin shell where the resonance condition $\omega_\mathrm{pl}(R'_\mathrm{res}) = m$ is satisfied, and detailed balance is achieved at the Bose-Einstein occupation number for the temperature of this shell: $T^\mathrm{res}_\odot= T(R'_\mathrm{res})$ (dashed gold curve). For $0.1\,\mathrm{eV} \lesssim m \lesssim 3\,\mathrm{eV}$, the resonance is (fractionally) least narrow and occurs in an outer Solar layer, 
so the hotter, denser layers interior to the resonant shell also contribute significantly, raising the effective temperature above $T_\odot^\mathrm{res}$. The occupation number of bound dark photon modes remains below that of equivalent-energy photons in the Solar core (at temperature $T_\odot^\mathrm{core}$, gray dashed curve), because the cooler outer layers of the Sun also contribute to emission and absorption.} \label{fig:fsat}
\end{figure}

In the previous section, we calculated the rate at which dark photons
are produced inside the Sun and emitted into bound orbits. However, particles will not remain on such orbits forever; they may decay, be reabsorbed by the Sun, or be gravitationally perturbed onto a different orbit, potentially even an unbound orbit.

\subsection{Decays}

The dominant radiative decay channel of a dark photon with mass $m<2m_e$ is to three SM photons ($A' \rightarrow 3A$), which in free space occurs at a rate~\cite{McDermott_2018}:
\begin{align}
	\tau_\mathrm{rad}^{-1} &\simeq 
	\frac{17 \epsilon^2 \alpha^4}{11664000 \pi^3} \frac{m^9}{m_e^8} \sim 10^{-40} {\rm \, y^{-1}} \left(\frac{\epsilon}{10^{-14}}\right)^2
	\left(\frac{m}{\keV}\right)^9.
\end{align}
(Decays to two photons are forbidden by the Landau-Yang theorem.) This estimate indicates that the decay rate for dark photons with $m \ll m_e$ is negligibly small over the age of the Solar System. This radiative channel is also far too inefficient to lead to an observable photon flux in e.g.~X-ray observatories.

In contrast, spin-0 particles such as axions can typically decay to two photons ($a \rightarrow 2 A$) through a dimension-5 operator such as $\LL \supset - \frac{1}{4}
g_{a \gamma\gamma} a F_{\mu\nu} \tilde F^{\mu\nu}$, leading to a radiative decay rate $\tau_\mathrm{rad}^{-1} \simeq {g_{a\gamma\gamma}^2 m_a^3}/{64 \pi} \approx 2 \times 10^{-9}\,\mathrm{y}^{-1}$ (for the largest allowed value of the axion-photon coupling $g_{a \gamma \gamma} = 10^{-10} \GeV^{-1}$ and $m_a = 1\,\mathrm{keV}$) that can be significant over the age of the Solar System. These axion decays could be an interesting signature for indirect detection of a Solar axion basin~\cite{DiLella_2003,tilburg2020stellar}.
The three-photon decay of the dark photon occurs through a dimension-8 operator, and is automatically very slow over the parameter space of interest.

\subsection{Absorption}

Reabsorption by the Sun can be an important effect.
From Eq.~\ref{eqdotf}, the rate of change of phase space density, at a point inside the Sun, is set by
\begin{equation}
	\dot f \simeq \Gamma_{\rm prod}' \left[1 + \left(1 - e^{m/T}\right) f\right]
\end{equation}
When production and reabsorption are dominated by the thin shell for which the $\omega_p \simeq m$ resonance is achieved, the occupation number $f$ for which there is no net change is $f_{\rm sat} \simeq {1}/({e^{m/T_\odot^\mathrm{res}}-1})$, i.e.~simply the bosonic thermal occupation number for the temperature at the resonance radius, $T_\odot^\mathrm{res}$.

In the simplest case, where particles always remain on the orbits
into which they were emitted, the maximum bound state density
at Earth would be given by $\rho_{\rm b}^{\mathrm{sat},i} = \frac{1}{(2 \pi)^3} m^4 f_{\rm sat}
\int_{\tilde V_i} \dd^3 v$, where $\tilde V_i$ represents the velocity phase
space volume of the initially occupied orbits.
For an Earth-crossing particle to have its perihelion closer than
some 
radius $R'$ inside
the Sun, its eccentricity must be very nearly unity, and its
angular momentum must be smaller than $R' \sqrt{2 |\Phi(R')|}$.
As a result, for emission from a shell of radius $R'$, we have
\begin{equation}
	\rho^{\mathrm{sat},i}_{\rm b,\odot} \simeq m^4 f_{\rm sat} \frac{1}{2 \pi^2} \frac{R'^2}{R^2} |\Phi(R')| v_{\rm esc}(R)
	\label{eqsatp}
\end{equation}
where $R = 1\, \mathrm{AU}$, and 
$v_{\rm esc}(R) \approx 1.4 \times 10^{-4}$ is the escape velocity at radius $R = 1\,\mathrm{AU}$.

However, as we will discuss in the next subsection,
Earth and other planets will gravitationally perturb
the orbits of the emitted particles. On timescales
short compared to the lifetime of the Solar
system, initially-Sun-crossing dark photon orbits
are perturbed to become non-Sun-crossing. Over time, the accessible velocity phase space volume will therefore grow larger
than $\tilde V_i$, and so will the corresponding saturation density at Earth. By Liouville's theorem, the phase space density is everywhere bounded by $f_{\rm sat}$,
so the maximum value for the bound state density at
Earth is:
\begin{equation}
	\rho^{\mathrm{sat}}_{\mathrm{b},\odot} = \frac{1}{(2 \pi)^3} m^4 f_{\rm sat}
	\int_{v < v_{\rm esc}} \dd^3 v
	= m^4 f_{\rm sat} \frac{v_{\rm esc}^3(R)}{6 \pi^2}.
	\label{rhosatfull}
\end{equation}
This is $\frac{2 G M R}{3 R'^2 |\Phi(R')|}$ times larger
than the ``unmixed'' initial saturation density from Eq.~\ref{eqsatp}. If we take $R'$ to be the mean emission radius (typically close to the radius at which resonance is achieved), this ratio is a factor of a few hundred throughout most of the Sun's volume.

Most of the above discussion assumed that only the parts
of the Sun at $T_\odot^{\rm res}(m)$ contribute significantly
to emission and reabsorption. More generally,
to have an equilibrium, we need
\begin{equation}
	\int \dd^3 R' \int \dd\tilde E  \,\frac{\dd Q}{\dd \tilde E} \left[1 + \left(1 - e^{m/T}\right) \bar f(R',\tilde E)\right]
	= 0. \label{eq:fsat1}
\end{equation}
where $\bar f(R',\tilde E)$ is the direction-averaged
phase space density at $R'$,
as a function of $\tilde E \equiv \tilde{\omega}_k + \Phi$, the \emph{total} energy per unit mass (negative for bound orbits). Gravitational perturbations are generally rather
slow to change $\tilde E$~\cite{tilburg2020stellar}---in particular, orbits that become
Earth-crossing almost always start out with $|\tilde E| \ll |\Phi(R')|$. We can therefore use $\frac{dQ}{d\tilde E} 
\simeq \widetilde Q(R')
\sqrt{|\Phi(R')|}$ in Eq.~\ref{eq:fsat1},
making it independent of $\tilde E$:
\begin{equation}
	\int \dd^3 R' \, \widetilde Q(R') \sqrt{|\Phi(R')|} \left[1 + \left(1 - e^{m/T}\right) \bar f(R')\right]
	\simeq 0.
\end{equation}
The simplest solution to this equation
is $\bar f(R') = (e^{m/T(R')}-1)^{-1}$, which corresponds
to local thermal equilibrium at all radii. This is the physical
solution when the dark photon's coupling is large
enough that its optical depth is small compared
to the Solar radius. However, for the unconstrained
parameter space we are interested in, the optical depth
is always large.
In this regime, the maximum basin density is obtained if mixing
is fast enough to make $f$ constant over the whole phase
space. In that case, $\bar f = f_{\rm sat}$ is constant,
with solution
\begin{equation}
	f_{\rm sat} \simeq \frac{\int \dd^3 R' \, \widetilde Q \sqrt{|\Phi|}}{\int \dd^3 R' \, \widetilde Q
	\sqrt{|\Phi|}	\left(e^{m/T} - 1\right)}. \label{eq:fsat2}
\end{equation}
Figure~\ref{fig:fsat} shows this $f_{\rm sat}$
as a function of dark photon mass. For masses which match
the Solar plasma frequency at some radius not too close to the surface,
the thermal occupation number at the resonant temperature
is a very good approximation. At lower masses, $f_{\rm sat}$
is somewhat higher than a naive extrapolation of this value,
since the higher-temperature interior parts of the 
Sun contribute significantly to the volume integrals.

The thick red line in Fig.~\ref{fig:rho} shows
the maximum bound state density from Eq.~\ref{rhosatfull},
using the $f_{\rm sat}$ value from Eq.~\ref{eq:fsat2} and Fig.~\ref{fig:fsat}.
This can be compared to the thin
red line, which shows the ``conservative'' saturated basin density from Eq.~\ref{eqsatp}, corresponding to
particles remaining on their initial orbits.

In Ref.~\cite{DiLella_2003}, the decay
of Solar basin axions within the Solar corona 
was proposed as a potential explanation for the
corona's high temperature, which is currently
somewhat mysterious~\cite{SAKURAI_2017}.
For the dark photon case, the reabsorption
rate scales with the plasma
density (or with the density squared, for bremsstrahlung).
The very low density of the Solar corona means
that heating from reabsorption there
is completely negligible, for all of the parameter
space that we will consider.

\subsection{Orbital perturbations}

In the previous subsection, we noted that gravitational
perturbations from Solar system bodies other than
the Sun can alter the orbits of the emitted dark photons.
One crucial question is how perturbations change
$\tilde E$, and in particular,
on what timescale orbits become unbound, i.e.\ $\tilde E > 0$.

To construct a simplistic model of this process,
we can assume that particles remain on their initial orbits,
until they are suddenly ejected after some time $t_{\rm eject}$,
which is exponentially distributed, with mean lifetime $\tau_{\rm eject}$.
Assuming that the
dark photon production rate has been constant 
throughout the lifetime of the Sun, the present-day basin density at Earth is given by
\begin{align}
\rho_\mathrm{b} = \max \left\lbrace \rho_\mathrm{b}^\mathrm{sat}, \dot{\rho}_\mathrm{b} \tau \right\rbrace,
\end{align}
where we defined an effective basin time:
\begin{align}
\tau \equiv \chi \tau_\mathrm{eject} \left[1 - \mathrm{exp}\left( - \frac{\tau_\mathrm{SS}}{\tau_\mathrm{eject}}\right) \right], \label{eq:tau}
\end{align}
with $\tau_\mathrm{SS} \approx 4.6\times
10^9\,\mathrm{y}$ the age of the Solar System.
The parameter $\chi$ is a fudge factor,
accounting for all of the effects that we have neglected.
These include perturbations that can the phase space
distribution of the orbits, deviations from the sudden ejection
approximation, and variation of the Sun's properties
over its lifetime.

Over the Sun's past evolution along the Main Sequence,
it has fused hydrogen into helium in its core. To
maintain equilibrium, its central density and
temperature have increased~\cite{Feulner_2012}. The resulting luminosity
increase means that the Sun is now about $30\%$
more luminous than it was shortly after its
formation. A careful treatment of dark photon
emission and absorption would take these changes in Solar
structure into account. However, since their
effects would most likely be at the $\OO(10\%)$ level,
we ignore them here.

A conservative scenario, leading to the lowest bound
density at Earth, is that planetary perturbations
can suddenly increase 
$\tilde E$
to $\ge 0$, and that this happens within a short
timescale, $\tau \simeq \tau_{\rm eject} \sim
10^7 \, \mathrm{y}$~\cite{tilburg2020stellar}. 
We can obtain a conservative value for
the saturation density by assuming that,
apart from ejections, gravitational perturbations do not
significantly change the initial particle orbits,
so the saturation density is approximately
given by Eq.~\ref{eqsatp}.

An optimistic scenario is
that gravitational ejection is so inefficient that
$\tau_{\rm eject} \gtrsim \tau_{\rm SS}$,
but mixing from gravitational perturbations is efficient enough to give
the saturation density from Eq.~\ref{rhosatfull}.
Correspondingly, we take our most optimistic
benchmark for the effective basin time to be $\tau 
= \tau_{\rm SS} \simeq 4.6 \times 10^9 {\rm \, y}$.
These ``conservative" and ``optimistic" benchmark timescales
are the same as those taken in Ref.~\cite{tilburg2020stellar}.

If perturbations can increase $\tilde E$, but do so gradually
rather than suddenly, then the bound state density at Earth
may be higher. This is because dark photons emitted from the Sun at lower
$\tilde E$ can become Earth-crossing, increasing
the phase space for emission that contributes to $\rho_\mathrm{b}(R)$. In effect, this could lead to $\chi > 1$ in Eq.~\ref{eq:tau}. On the other hand, an effect that slightly lowers $\chi$ is the departure from purely radial orbits (with eccentricity $e \simeq 1$) due to gravitational perturbations from the planets. This ``mixing" of eccentricities is likely significantly faster than gravitational ejection. The result of deforming all orbits to a fixed eccentricity $e$ is to effectively lower the RHS of Eq.~\ref{eq:rhodot} by a factor of $(2+e^2)/3$. A ``fully mixed'' eccentricity distribution $f(e) = 2e$ (between $0\leq e \leq 1$), corresponding to uniform phase space density, would lead to an average reduction factor of $5/6$ for the basin density.

The gravitational ejection time of particles without non-gravitational interactions has recently been studied in Ref.~\cite{anderson2020direct}, which implemented the phase-space diffusion dynamics of Ref.~\cite{gould1991gravitational} in the context of a primordial DM abundance bound to the proto-Solar molecular cloud. It was found that 31\% of Earth-crossing density (excluding Jupiter-crossing orbits with $a > a_\mathrm{J}/2$, which were assumed to be quickly ejected) survived the age of the Solar System, implying a gravitational ejection time of $\tau_\mathrm{eject} \approx 4.0 \times 10^9 \,\mathrm{y}$ with their initial conditions. The phase space into which basin dark photons are emitted is likely slightly longer lived, due to the steeper distribution of semi-major axes. The injected fraction of Earth-crossing basin energy density that can also cross Jupiter's orbit is only $\simeq (16/15\pi) (\mathrm{AU}/a_\mathrm{J})^{5/2}$; since $a_\mathrm{J} \approx 5.2 {\rm \, AU}$, this is a sub-1\% correction, which we ignore here. Taking $\chi = 5/6$ to account for eccentricity redistribution, we arrive at a fiducial estimate of $\tau \approx 2.3 \times 10^9\,\mathrm{y}$ based on the results of Ref.~\cite{anderson2020direct}. We refrain from using this estimate to place lower limits on $\rho_\mathrm{b}$ until a full orbital dynamics simulation (along the lines of e.g.~Ref.~\cite{Peter_2009,peter2009dark}) is performed that includes all the planetary orbits without resorting to time-averaging, and that uses our peculiar radial-orbit initial conditions.

The appropriate saturation density could also be obtained
from such simulations. It is clear that, strictly speaking,
particles do not remain on their initial orbits.
Even ignoring planetary perturbations, the 
gravitational potential deviates from $\Phi \propto -1/R$ within the Solar radius, so
orbits with perihelion inside the Sun will precess.
However, since emission from the Sun is isotropic,
different orientations are populated equally, so this
precession does not change the phase space distribution.
We are interested in processes which can change
the action variables for the orbit~\cite{binney2011galactic},
rather than just the angle variables.

The strongest perturbing effects are often those of Jupiter.
If a particle's orbit is Jupiter-crossing, gravitational interactions will eject
the particle from the Solar System on timescales much shorter than $\tau_{\rm SS}$~\cite{gould1991gravitational}.
For non-Sun-crossing particles with lower energies, Jupiter's perturbations cause
orbits to undergo ``Kozai cycles", which
approximately conserve the semi-major axis $a$ and angular momentum component $L_z$ perpendicular to the ecliptic plane, but change 
the orbit's eccentricity~\cite{Murray_2000}.
This opens up more phase space volume than the initial $\tilde{V}_i$ (with $e \simeq 1$
orbits), though since eccentricity and orbital inclination
are related by the conservation of $L_z$, it does not
explore the full phase space.
An important question is whether other effects, such
as the eccentricity of Jupiter's orbit~\cite{Katz_2011},
or perturbations from other planets~\cite{gould1991gravitational,Peter_2009}, allow particles to explore the full phase space,
and on what timescale.

An additional difficulty, for particles emitted deep inside
the Sun, is escape to non-Sun-crossing orbits.
Due to the rapid perihelion precession noted
above, the perturbing torques from Jupiter (and other planets)
average out as the orbit precesses around~\cite{Damour_1999}.
In Ref.~\cite{Peter_2009}, it was estimated that
close encounters with Earth and Venus are the main effect
raising the perihelion of these orbits,
and can raise them out of the Sun over a timescale of order $10^8 {\rm \, y}$. A more detailed analysis of this process would be valuable.

In summary, particles emitted from inside the Sun will eventually populate the entire velocity phase space, but detailed simulations of the dynamics of this phase space migration are needed. For our fiducial case, we take
the simple, optimistic assumption that $\OO(1)$ of the full
phase space (for fixed $a$) is explored.
This gives the saturation density from Eq.~\ref{rhosatfull}.
A more conservative assumption would be that only
a subspace with conserved $L_z$ is explored, as
is the case for Kozai cycles. This would give
\begin{equation}
	\rho_{\rm b}^{\mathrm{sat},K} \simeq m^4 f_{\rm sat} \frac{1}{4\pi^2}
	\frac{R'}{R} \sqrt{2 |\Phi(R')|} v_{\rm esc}(R)^2
\end{equation}
in the notation of Eq.~\ref{eqsatp}, which is a factor
$\frac{3}{2} \frac{R'}{R} \frac{\sqrt{2 |\Phi(R')|}}{v_{\rm esc}(R)}$ smaller than the full saturation density from
Eq.~\ref{rhosatfull}. Throughout most of the Sun,
this ratio is $\OO(10)$.
This gives a rough estimate of how much smaller the saturation density might be,
though we emphasize once again that a simulation-based analysis would be required to
determine these quantities with confidence.

\subsection{Bound state density}

\begin{figure*}
\includegraphics[width = 0.8 \textwidth]{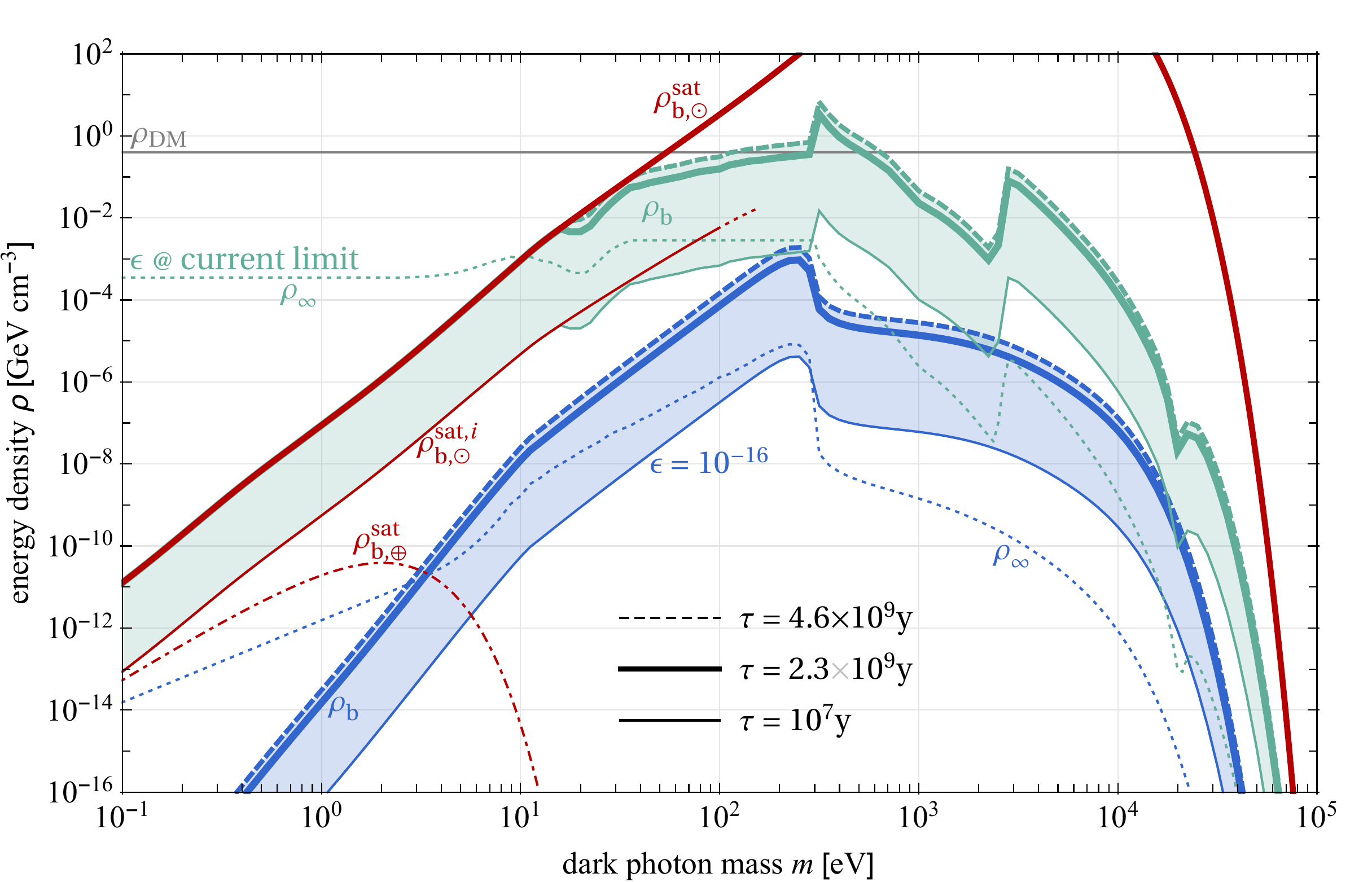}
\caption{Energy density $\rho$ of dark photons with mass $m$ at Earth's location ($R = 1\,\mathrm{AU}$) in the Solar System. Blue curves assume a fixed kinetic mixing parameter $\epsilon = 10^{-16}$; they represent the Solar dark photon basin energy density $\rho_\mathrm{b}$ for three values of gravitational ejection times ($\tau = 10^7\,\mathrm{y}$, thin; $\tau = 2.3 \times 10^9 \,\mathrm{y}$, thick; $\tau = 4.6\times 10^9\,\mathrm{y}$, dashed), as well as the unbound, relativistic emission $\rho_\infty$ (thin dotted). The bound dark photons dominate the unbound relativistic flux in terms of energy density ($\rho_\mathrm{b} > \rho_\infty$) for $m \gtrsim 0.3\,\mathrm{keV}$, and possibly as low as $m \gtrsim 2.5\,\mathrm{eV}$ for the optimistic $\tau = 4.6 \times 10^9\,\mathrm{y}$. Green curves represent equivalent quantities for $\epsilon$ at the current limit, i.e.~the lower boundary of the gray shaded region in Fig.~\ref{fig:epsilon}. At low masses $m \lesssim 15\,\mathrm{eV}$, the Solar basin may achieve detailed balance with the Sun, and approach the initial Solar-basin saturation density $\rho_\mathrm{b,\odot}^{\mathrm{sat},i}$ (thin red) of Eq.~\ref{eqsatp} with the conservative value of $\tau$, or the full saturation density $\rho_\mathrm{b,\odot}^{\mathrm{sat}}$ of Eq.~\ref{rhosatfull} (thick red) with fiducial and optimistic values of $\tau$. The saturation density $\rho_{\mathrm{b},\oplus}^\mathrm{sat}$ of Earth's basin is below the thin dot-dashed red curve. At higher masses, the Solar basin can be an appreciable fraction of the average dark matter energy density $\rho_\mathrm{DM}$ (gray); the area where $\rho_\mathrm{b} \gtrsim \rho_\mathrm{DM}$ is now excluded~(cfr.~Fig.~\ref{fig:epsilon}).} \label{fig:rho}
\end{figure*}

Putting all of these pieces together,
Fig.~\ref{fig:rho} illustrates the bound state
densities for the aforementioned scenarios.
The blue curves show the basin densities that would arise from
a fixed, small value of $\epsilon$, as a function of dark photon mass. Even for the most conservative gravitational ejection time $\tau = 10^7 \, \mathrm{y}$, there is a wide
mass range in which the bound density dominates the density due to the
unbound flux. For the optimistic scenario of $\tau = 4.6 \times 10^9 \, \mathrm{y}$, the basin energy density exceeds the unbound energy density all the way down to $m \sim 3\, \mathrm{eV}$.

The green curves show the bound densities that can be attained
at the largest possible $\epsilon$, consistent with previous constraints
(shown as the gray region in Fig.~\ref{fig:epsilon}).
For $m \lesssim 15 \eV$, these may reach the saturation densities discussed
above. At masses of order a few hundred electronvolt, densities close to the Galactic DM density
of $\rho_{\rm DM} \simeq 0.4 \, \mathrm{GeV}\,\mathrm{cm}^{-3}$ can be obtained.

Dark photons can also be produced in other hot, dense Solar System environments. The Earth's core
and mantle are of particular interest, since particles sourced
there can accumulate in Earth's gravity well. However, 
the relatively low temperatures inside the Earth
($T \lesssim 0.5\,\mathrm{eV}$), and the high plasma frequency in its
iron core, mean that the emission rate will not be very large
(though a precise calculation is not trivial).
Furthermore, the small escape velocity at the Earth's surface,
$v_{\rm esc}^\oplus \simeq 11 {\rm \, km \, s^{-1}}$, means
that the saturation density is fairly small.
Since the maximum temperature anywhere inside
the Earth is $T_{\max} \sim 6000 \, \mathrm{K}$~\cite{Anzellini_2013},
the saturation phase space density for bound orbits is
$f_{\rm sat} \le f_{\max} = (e^{m/T_{\max}}-1)^{-1}$.
Consequently, the saturation density at the Earth's surface
is bounded from above by
\begin{equation}
	\rho_{\rm b,\oplus}^{\mathrm{sat}} \le m^4 f_{\max}(m) \frac{(v_{\rm esc}^{\oplus})^3}{6\pi^2}
\end{equation}
This bound is plotted in Fig.~\ref{fig:rho}, which shows
that it is always significantly smaller than the Solar
basin saturation density at Earth. In addition,
it is also smaller than the conservative basin density
sourced by the Sun, at kinetic mixings that will be relevant for
experiments (discussed in the next section). We conclude that the Solar basin densities will be dominant for all practical purposes.

\section{Direct detection}

A Solar basin of dark photons can be directly detected in the laboratory by the same experiments that look for dark-matter dark photons. Both populations have very small velocity dispersions and mean velocities in Earth's reference frame---$v \sim 10^{-3}$ for DM and $v \sim 10^{-4}$ for the Solar basin. All existing experiments~\cite{akerib2017first,fu2017limits,Wang_2020, Aralis_2020,abe2018search, armengaud2018searches,aprile2017search,aprile2019light,aprile2020observation} and practically feasible proposals are based on absorption of the dark photon's rest-mass energy, where the detector's capabilities are usually not sufficient to resolve the spread in kinetic energy or to provide directional information. (Exceptions include Ref.~\cite{Arvanitaki_2018}, a proposal to use narrow absorption lines of molecules in the gas phase for dark photons with $m \lesssim 20\,\mathrm{eV}$, and coherent absorption proposals
such as those in~\cite{Baryakhtar_2018,Arvanitaki_2018}, where the
angular spread of the absorbed photons corresponds to the
dark photon velocity distribution.)

In this limit of negligible velocities, the absorption rate of a nonrelativistic dark photon of mass $m$ is solely determined by the coupling $\epsilon$ and the ambient density $\rho$ (be that of DM or the Solar basin), in the combination $\epsilon^2 \rho$. For the Solar basin, the ambient density $\rho_\mathrm{b}$ is proportional to another factor $\epsilon^2$, unless $\epsilon$ is large enough to achieve the saturation level $\rho_\mathrm{b} \simeq \rho_\mathrm{sat}$ (e.g.~Eq.~\ref{rhosatfull}, for the case of a fully mixed phase space). As in Ref.~\cite{tilburg2020stellar}, we can thus recast any DM limit on (or prospective sensitivity to) the kinetic mixing $\epsilon_\mathrm{DM}$ into the equivalent quantity $\epsilon_\mathrm{basin}$ for dark photons in the Solar basin, via the map:
\begin{align}
\hspace{-0.3em} \epsilon_\mathrm{basin} = \max \hspace{-0em} \left\lbrace \sqrt{\epsilon_\mathrm{DM}}\left[\frac{\rho_\mathrm{DM}}{\dot{\rho}_\mathrm{b}\big|_{\epsilon = 1} \tau} \right]^{1/4}, \epsilon_\mathrm{DM} \sqrt{ \frac{\rho_\mathrm{DM}}{\rho^{\mathrm{sat}}_{\mathrm{b},\odot}} }\right\rbrace, \label{eq:recast}
\end{align}
with $\dot{\rho}_\mathrm{b}\big|_{\epsilon = 1}$ the basin energy density injection rate at $\epsilon = 1$. For the conservative limit with $\tau = 10^7\,\mathrm{y}$, we take in Eq.~\ref{eq:recast} the saturation density to be the lower $\rho_{\mathrm{b},\odot}^{\mathrm{sat},i}$ of Eq.~\ref{eqsatp} instead of $\rho_{\mathrm{b},\odot}^{\mathrm{sat}}$ of Eq.~\ref{rhosatfull}.
The $\epsilon_\mathrm{basin}$ limit or discovery reach will generally be weaker than $\epsilon_\mathrm{DM}$, in view of the lower basin density $\rho_\mathrm{b} < \rho_\mathrm{DM}$ in most of parameter space (cf.~Fig.~\ref{fig:rho}). However, the basin bound on the kinetic mixing does \emph{not} require the assumption that dark photons constitute all or even part of the dark matter energy density, and is thus more general (like the stellar cooling constraints on $\epsilon$).

\begin{figure*}
\includegraphics[trim = 0 0 0 22, clip, width = 0.89 \textwidth]{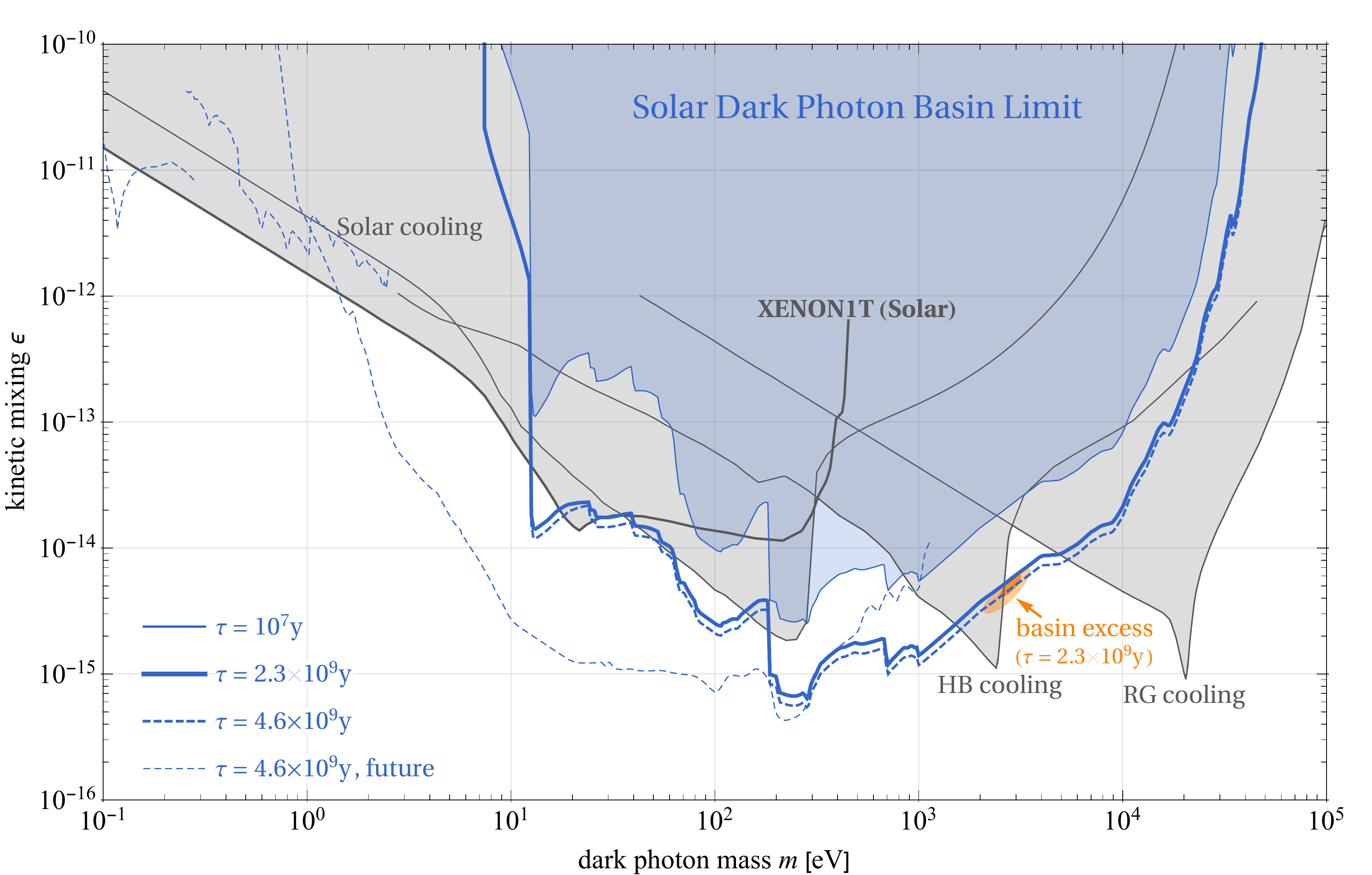}\\
\vspace{0em}
\includegraphics[trim = 0 0 0 22, clip, width = 0.89 \textwidth]{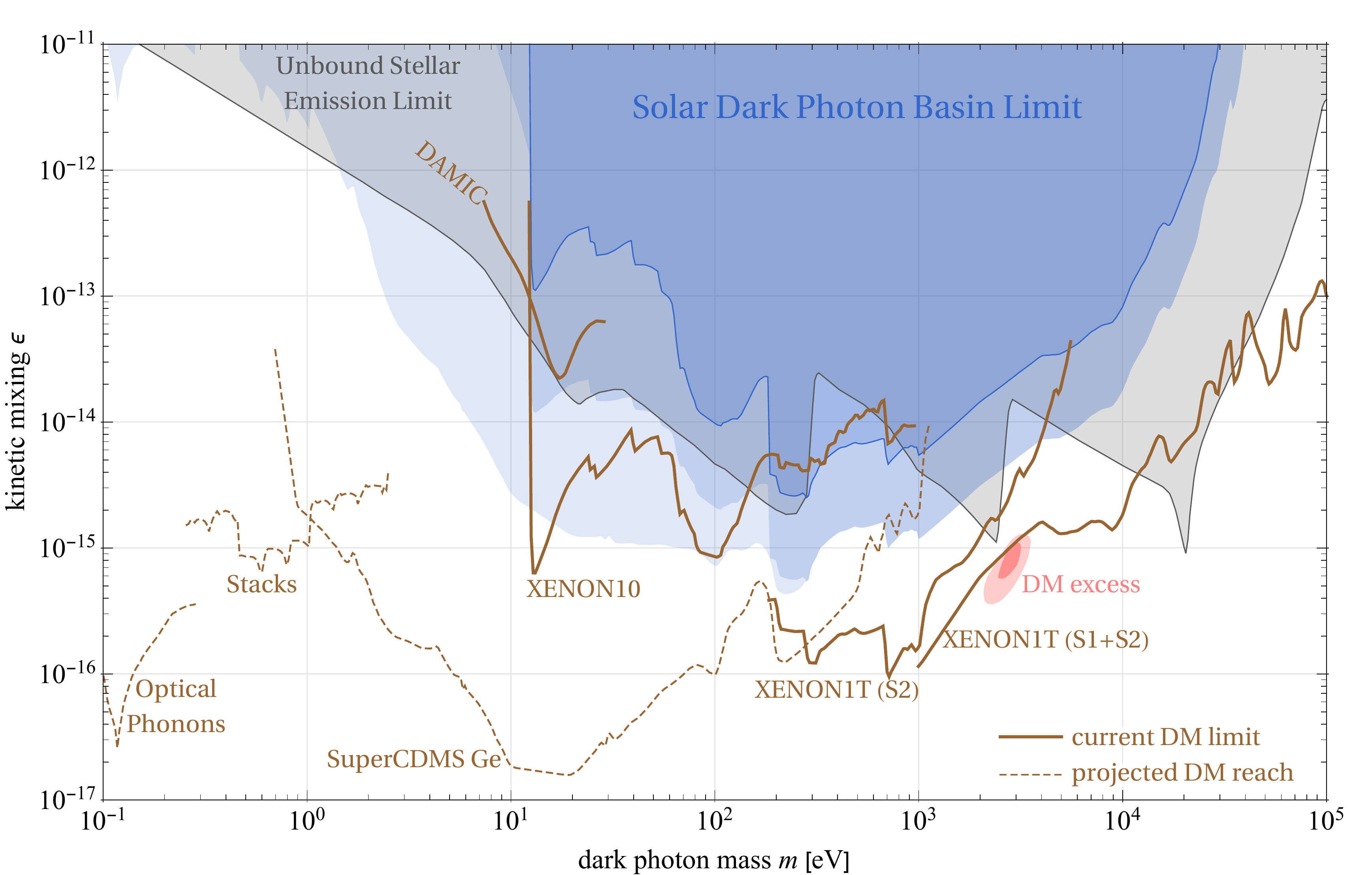}
	\caption{Constraints and projected reach for the kinetic mixing $\epsilon$ of a dark photon with mass $m$. \textit{Top panel:} The shaded blue region is a new Solar basin limit from a recast of dark photon line absorption searches~\cite{Aguilar_Arevalo_2017,Angle_2011,Bloch_2017,aprile2019light,aprile2020observation} using Eq.~\ref{eq:recast} with a conservative gravitational ejection time estimate $\tau = 10^7\,\mathrm{y}$. Thick and dashed blue lines delineate parameter space that is disfavored (fiducial $\tau = 2.3 \times 10^9 \,\mathrm{y}$) and where signals would have been plausible (optimistic $\tau = 4.6 \times 10^9\, \mathrm{y}$). Thin dashed curves depict the potential discovery reach to basin dark photons with selected future experiments~\cite{Knapen_2018,Griffin_2018,Baryakhtar_2018,Bloch_2017,GolwalaTalk}. The orange 1- and 2-$\sigma$ error ellipses are best-fit parameters of XENON1T's excess~\cite{aprile2020observation} interpreted as a dark-photon Solar basin with $\tau = 2.3 \times 10^9\,\mathrm{y}$. The gray area is excluded by a combination of the unbound Solar dark photon search of Ref.~\cite{an2020new} and absence of anomalous cooling of the Sun~\cite{An_2013,Redondo_2013}, horizontal branch (HB)~\cite{Giannotti_2016}, and red giant (RG)~\cite{viaux2013neutrino} stars. \textit{Bottom panel:} Dark-photon dark matter limits of four  leading-sensitivity analyses (thick brown) and projected sensitivity of three representative future experiments (dashed brown). Pink error ellipses are best-fit DM parameters of excess in XENON1T. These DM absorption limits and projections are used for the recast basin limit, region of interest, and future projection, shaded in blue in this panel.} \label{fig:epsilon}
\end{figure*}

In Fig.~\ref{fig:epsilon}, we recast
upper limits $\epsilon_\mathrm{DM}$
from dark-photon dark matter searches
by DAMIC~\cite{Aguilar_Arevalo_2017},
XENON10~\cite{Angle_2011,Bloch_2017}, and
XENON1T~\cite{aprile2019light,aprile2020observation}, shown as thick brown curves in the bottom
panel, into a Solar dark photon basin limit
$\epsilon_\mathrm{basin}$ (shaded blue region in
top panel, darkest shade of blue in bottom panel)
using the conservative gravitational ejection time
of $\tau = 10^7\,\mathrm{y}$. The true value for
$\tau$ is likely significantly higher, as argued
above and in Ref.~\cite{tilburg2020stellar}, with
the fiducial value $\tau = 2.3 \times 10^9\,\mathrm{y}$
disfavoring the parameter space above
the thick blue line in the top panel of
Fig.~\ref{fig:epsilon}. 
This is close to the `optimistic' estimate of 
$\tau = \tau_{\rm SS} \simeq 4.6 \times 10^9 {\rm \, y}$,
indicated by the dashed blue curve of 
Fig.~\ref{fig:epsilon}'s top panel (and by
the intermediate shade of blue in the bottom panel).
The tentative excess in the recent
XENON1T analysis~\cite{aprile2020observation}
could be interpreted in terms of a DM dark
photon (pink 1- and 2-sigma error ellipses in
Fig.~\ref{fig:epsilon})~\cite{an2020new}, but
could also be due to a Solar-basin dark photon
(orange error ellipses, assuming the fiducial value
$\tau = 2.3 \times 10^9\,\mathrm{y}$).

Future DM experimental proposals for
low-mass dark photons are shown as thin dashed
brown curves in the bottom panel. We pick three
representative sensitivity projections covering
a broad range of dark photon masses; 
polar crystals in which DM can convert to optical
phonons~\cite{Knapen_2018,Griffin_2018},
dielectric haloscopes constructed from layered
stacks of refractive materials~\cite{Baryakhtar_2018} 
(we take the experimental parameters from~\cite{Lasenby:2019hfz}),
and semiconductor targets, represented by projections
for SuperCDMS~\cite{Bloch_2017,GolwalaTalk}.
We show that these proposals are also sensitive to
Solar basin dark photons, potentially probing new
parameter space down to $m \sim 0.1\,\mathrm{eV}$,
as indicated by the thin dashed blue curves in
the top panel (lightest blue shaded region in the
bottom panel), assuming the optimistic value $\tau
= 4.6 \times 10^9 \, \mathrm{y}$.

The unbound Solar flux of dark photons peaks
at energies $\sim 10 \eV$ (for $m \lesssim 10 \eV$)
\cite{Redondo_2013}, so experiments 
with sensitivity at these energies, such as SuperCDMS,
may also improve the constraints from (or detect) unbound emission.
However, as indicated in Figure~\ref{fig:rho},
the bound density can exceed the unbound density for 
$m \gtrsim 3 \eV$, and it can have a higher number density
at even lower masses. Moreover, the bound population results
in an almost-monoenergetic absorption signal,
while the unbound flux is over an $\OO(1)$ spread in energies,
so the former may be easier to distinguish from backgrounds.

The combined constraints from the unbound dark photon emission are shown collectively as the gray region; from left to right, a relativistic Solar dark photon search in XENON1T~\cite{an2020new}, the parameter space in which the Solar luminosity would be changed by 10\% or more~\cite{An_2013,Redondo_2013}, and horizontal-branch~\cite{dearborn1990dark} and red-giant~\cite{viaux2013neutrino} stellar cooling constraints taken from Ref.~\cite{An_2013}.

In summary, our Solar basin re-interpretation of the dark matter limit of Ref.~\cite{aprile2019light} robustly excludes new parameter space for dark photon masses $0.27\,\mathrm{keV } \lesssim m \lesssim 0.83\,\mathrm{keV}$, improving by up to an order of magnitude over the previously most stringent DM-independent bounds from Solar and HB-star cooling. With larger values of the gravitational ejection timescale, a large region of parameter space between $12\,\mathrm{eV}$ and $5.4\,\mathrm{keV}$ is already being probed by current liquid-xenon experiments, again independent of assumptions about DM. Next-generation versions~\cite{Aprile_2016,LZ,Aalbers_2016} are set to provide another near-future leap in sensitivity. At the low-mass end, innovative low-threshold experimental proposals may extend the potential discovery reach to dark photon masses as low as $m\sim 0.1\,\mathrm{eV}$.

The Solar-basin versus dark-matter interpretations of any putative positive signal can be disentangled (with sufficient statistics) by their differences in annual modulation. With a $1/R^4$ scaling of the basin energy density, Earth's eccentric orbit would lead to a fractional annual modulation amplitude of 6.7\% of the density, with the peak signal reached at perihelion $R = 0.9833\,\mathrm{AU}$ on January 3, and significantly lower signal at aphelion $R = 1.0167\,\mathrm{AU}$ six months later. (For a fully mixed saturated basin, the density scales like $1/R^{3/2}$, giving a 2.5\% modulation amplitude.) The basin energy density may also be temporally intermittent due to the chaotic nature of the orbits in the presence of planetary perturbations, which should be a topic of future study. In contrast, the DM energy density would exhibit much weaker annual modulation with fractional amplitude of about 1.5\% due to gravitational focusing~\cite{Lee_2014}, with peak signal achieved around March 1, when Earth is ``downwind'' from the Sun in the DM halo (in the laboratory frame). 


\section{Discussion}

In this work, we have shown that the Solar dark photon
basin---the nonrelativistic dark photons
emitted by the Sun into bound orbits---lends itself to
direct detection in the laboratory over a large range of
masses, from $0.1\,\mathrm{eV}$ to $10^4\,\mathrm{eV}$. 
The importance of collective effects in dark photon emission, leading to resonant emission and low-mass decoupling, 
are responsible for the larger observable parameter space (not already excluded by stellar
constraints) compared to the case of axions coupled to
electrons studied in Ref.~\cite{tilburg2020stellar}, 
which first studied the absorption of Solar basin particles
in dark matter detection experiments.
We
showed that the Solar basin may even reach detailed balance
with the Solar emission volume at low dark photon masses and
near the current kinetic-mixing limit.

Without the assumption that dark photons make up
all (or any fraction of) the DM energy density,
existing searches (Ref.~\cite{aprile2019light}
in particular) for nonrelativistic dark photon
absorption have already (but unknowingly) set
world-leading DM-independent constraints on
the kinetic mixing parameter around $m \sim
0.5\,\mathrm{keV}$. It is possible that the excess
reported by Ref.~\cite{aprile2020observation}
is caused by absorption of $m \approx
2.8\,\mathrm{keV}$ dark photons in the Solar
basin. Future experiments will drastically extend
the discovery reach, in terms of kinetic mixing
and especially towards lower dark photon masses.

As noted already in
Ref.~\cite{tilburg2020stellar}, it is imperative
that simulations of orbital dynamics be carried
out to determine the lifetime, evolution, and
statistical behavior of the Solar basin of any
weakly coupled particle. Such studies would
drastically sharpen our predictions, and provide
critical input to data analyses of direct
detection experiments. Finally, we note that the
mechanism described in this work would also be
operative in other astrophysical contexts. Around
more extreme stars or compact remnants, the large
``dark electromagnetic field strengths'' may give
rise to interesting signatures in more complicated
models that include a dark photon.

\medskip

\acknowledgments{We thank Timothy Wiser for helpful discussions, and Maxim Pospelov and Diego Redigolo for comments on our manuscript. KVT's research is funded by the Gordon and Betty Moore Foundation through Grant GBMF7392, and supported in part by the National Science Foundation under Grant No.~NSF PHY-1748958.}

\bibliography{PhotonBasin}

\end{document}